# Analysis of the Arrival Directions of Ultrahigh Energy Cosmic Rays

A.A. Mikhailov

Yu.G. Shafer Institute of Cosmophysical Research and Aeronomy, 31 Lenin Ave., 677980 Yakutsk, Russia
mikhailov@ikfia.ysn.ru

**Abstract:** The arrival directions of ultrahigh energy extensive air showers (EAS) by Yakutsk, AGASA and SUGAR array data are considered. For the first time, the maps of equal exposition of celestial sphere for the distribution of particles by AGASA and SUGAR array data have been constructed. The large-scale anisotropy of $E>4\cdot10^{19}$ eV cosmic rays from the side of Input and Output of the Galaxy Local Arm by Yakutsk, AGASA and SUGAR array data has been detected. The problem of cosmic ray origin is discussed.

**Introduction**

Until there is an opinion, that cosmic rays with energy $E>4\cdot10^{19}$ eV are isotropic (see for example [1,2]). Here data of array EAS Yakutsk, AGASA, SUGAR with take into account their exposition on celestial sphere are analyzed.

**Experimental data and discussion**

At first, we have analyzed Yakutsk EAS array data whose shower cores lie inside the array perimeter and the accuracy to determine the arrival angle is ~ 3°. The particle energy is estimated by a new formula according to [3]. Fig.1 presents the distribution of 34 particles with $E>4\cdot10^{19}$ eV on the map of equal exposition of celestial sphere (the method to construct this map is based on the estimation of the expected number of showers [4,5] and etc.). At the map of equal exposition the equal number of particles from the equal parts of sphere is expected. As seen in Fig.1, the particles practically have isotropic distribution on the celestial sphere. However, the most concentration of particles is observed from the side of Input of Galaxy Local Arm at galactic latitude $3.3°<b<29.7°$ and longitude $60.1°<l<116.8°$ (this region it is noted by dash quadrangles). In this coordinates there are 9 particles. The probability of chance to find 9 of 34 particles in this coordinates by method the Monte Carlo [5] is P~0.014. If we decrease the considered range within $3.3°<b<19.1°$ and $60.1°<l<116.8°$ then the probability of chance to find 7 of 34 particles will be P~0.008.

For the distribution of particles with $E>4\cdot10^{19}$ eV by AGASA array data [6] we construct the map of equal exposition of celestial sphere according to [4,5]. As seen from this map, almost a half of events (25 particles of 58) are within of coordinates toward the side of Input the Local Arm $11.2°<b<69.3°$ and $38.9°<l<154°$. The probability of chance to find 25 particles of 58 at abovementioned coordinates is P~0.0004. If we decrease the limits of considered coordinate range up to $19.4°<b<29.8°$ and $63.5°<l<108.8°$, then the probability to find 7 particles of 58 will be P~0.0003. Thus, the statistically significant particle flux in the case of the AGASA array is observed from the side of the Input of Local Arm as data of Yakutsk EAS array.

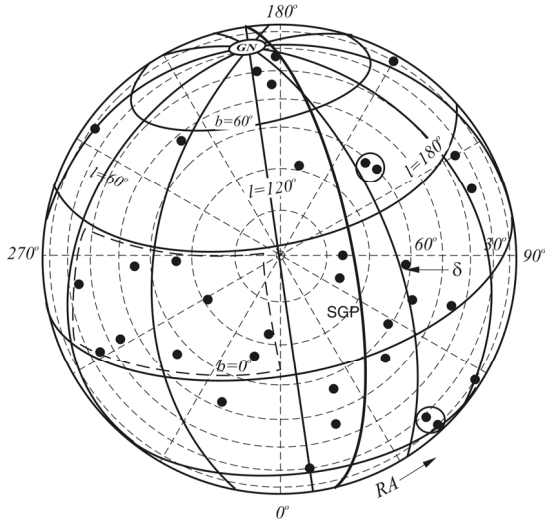
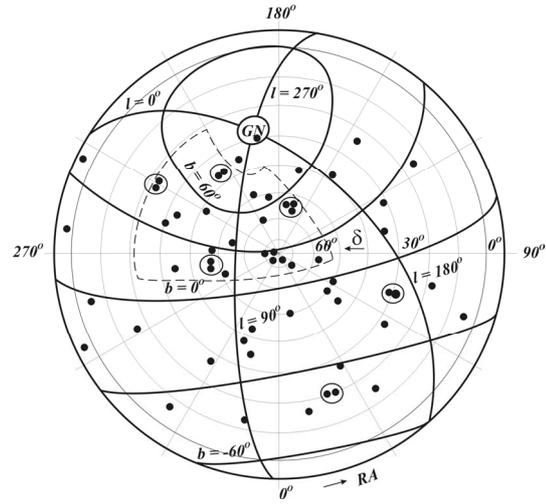

Fig.1. On the map of equal exposition particles with E>4.10$^{19}$ eV are shown by the Yakutsk EAS array data. SGP – Super Galactic Plane (plane of Local cluster). Dashed quadrangles on the left – a considered region of a celestial sphere. δ - declination, RA – right ascension, b, l – galactic latitude and longitude. Big circles – clusters.

Fig.2. The same as in Fig.1 for the AGASA array data. Dashed quadrangles on the upper a considered region of a celestial sphere.

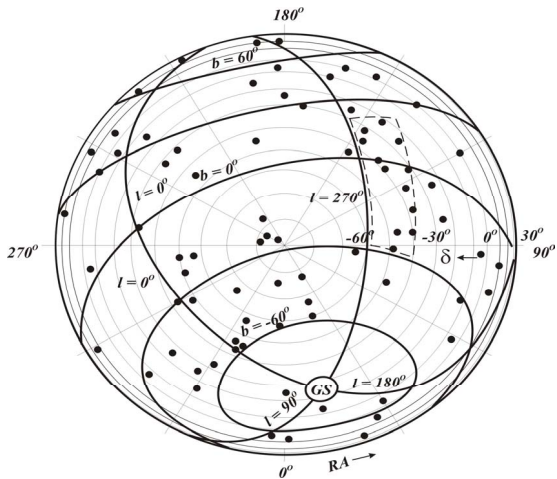

Fig.3. The same as in Fig.1 for the SUGAR array data. Dashed quadrangles on the right – a considered region of a celestial sphere.

For the distribution of particles with E>4.10$^{19}$ eV of the SUGAR array [7] the map of equal exposition of celestial sphere is also constructed. In [8] have shown, that model 'Hillas" of an estimation of energy of EAS is more correct, according to this model 80 particles with E>4.10$^{19}$ eV are registered.

The most concentration 12 of 80 particles [7] is observed toward the Output of Local Arm within coordinates -28.9°< b <19.3° and 248°<b<267.3° with the probability of chance P~2.10$^{-5}$.

We found [9] if the sources of particles are distributed in the Galaxy disc, then the protons with E~10$^{18}$ eV in main move along the field lines of the Galaxy Arms. On the basis, it may be suggested that the observed flux of particles with E>4.10$^{19}$ eV from the side of the Input and Output of Galaxy Local Arm has a rigidity $R \sim 10^{18}$ eV and therefore they are the charged superheavy particles. The similar conclusion was made by us on the basis of experimental data analysis on the distribution of particles in zenith angle of observations and in muon composition in extensive air showers at $E=10^{19}-10^{20}$ eV by Yakutsk EAS array data [8,10].

Note, that the increased flux of particles is observed from an Input of a Arm (from site of the center of the Galaxy) above a plane of the Galaxy, from an Output of a Arm (from site of the anticenter) – in main below a plane. Such distinction of flux of particles from the center and the anticenter of the Galaxy is expected in galactic model of an origin cosmic rays with mainly azimuthally of a large-scale magnetic field if sources of particles are in the disc of the Galaxy [11].

Early by data of arrays EAS Haverah Park and Yakutsk at $E \sim 10^{19}$ eV the increased flux of particles from the anticenter at latitude $b<0°$ were found [5,12]. We consider the arrival directions of EAS by data Yakutsk, AGASA, SUGAR with $E>4.10^{19}$ eV for the center in longitude $-90°<l<90°$ and for the anticenter $90°<l<270°$ at upper/below galactic plane of Galaxy. Total number of particles of 3 arrays is equal 172. In Fig.4a from the center of Galaxy it is shown the ratio number of particles with take into account exposure of arrays $R=n_1(b>0°)S_2/n_2(b<0°)S_1$, where $n_1$, $n_2$ –number of particles at $b>0°$ and $b<0°$, $S_1$ and $S_2$ are exposure of the celestial sphere to arrays at $b>0°$ and $b<0°$ according [5]. If the cosmic rays are isotropy then $R = 1$. According to Fig.4a the increased flux of particles is observed from the center of the Galaxy at latitude $b>0°$ and to Fig.4b the increased flux of particles - from the anticenter at latitude $b<0°$, as predict in galactic model of an origin cosmic rays [11].

It is note, that from northern and southern poles of the Earth at $\delta \sim \pm 90°$ by data of arrays AGASA and SUGAR (Fig.2, 3) concentration of density of distribution of particles is observed.

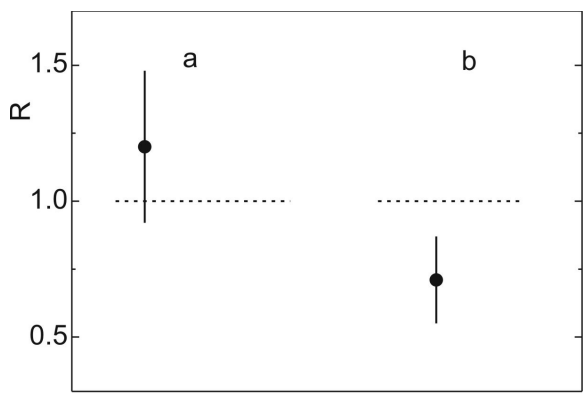

Fig.4. Ratio number of particles R with take into account exposure of arrays at b>0°/b<0°: a – from center of Galaxy, b - from the anticenter of Galaxy.

**Conclusion**

The large–scale anisotropy of particles at $E>4.10^{19}$ eV from the side of the Input and Output of Galaxy Local Arm has been found. The particles of ultrahigh energies are most likely the superheavy nuclei and they have a galactic origin.

**References**

[1] M. Nagano M., A.A. Watson. Review of Modern Physics. 72, 689, 2000.

[2] V.S. Berezinsky, A. Gazizov, S. Grigoreva. Physical Review, 74, 043005, 2006.

[3] M.I. Pravdin, A.V. Glushkov, A.A. Ivanov et al. Proc. 29-th ICRC, Pune, 7, 243, 2005.

[4] A.A. Mikhailov. Bulletin Nauchno-Technical informaion. Problemi Kosmophysiki i Aeronomii. Yakutsk, December, 9, 1982.

[5] N.N. Efimov, A.A. Mikhailov, M.I. Pravdin. Proc. 18-th ICRC, Bangalore, 2, 149, 1983.

[6] N. Hayashida, K. Honda, N. Inoue et al. Astrophys. J., 522, 225, 1999.

[7] M.M. Winn, J. Ulrichs, L.S. Peak et al., J. Phys. G.: Nucl. Phys., 12, 653, 1986.

[8] A.A. Mikhailov. Pisma v Zhetf., 79, 175, 2004.

[9] V.S. Berezinsky, A.A. Mikhailov Proc. 18-th ICRC, Bangalore, 2, 174, 1983.

[10] A. A. Mikhailov, N.N. Efremov, N.S. Gerasimova et al. Proc. 29-th ICRC, Pune, 7, 227, 2005.

[11] S.I. Syrovatsky. Preprint of P. N. Lebedev Physical institute of AS USSR, M., 7, 1969.

[12] S.M. Astley, G. Gunningham G., J. Lloyd-Evans et al. Proc.17-th ICRC, Paris, 2, 156, 1981.